\documentclass[prb,twocolumn,amssymb]{revtex4}
\usepackage{amssymb}
\usepackage{amsmath}
\usepackage{graphicx}

\newcommand{\la}{\langle}
\newcommand{\ra}{\rangle}

\begin{document}

\title{Quantum Antiferromagnetism of Fermions in Optical Lattices with Half-filled $p$-band }
\author{Kai Wu$^{1}$ and Hui Zhai$^{1,2,3}$} \affiliation { 1) Center for Advanced Study, Tsinghua University, Beijing, 100084, China\\ 2) Department of Physics, University of California,
Berkeley, California, 94720, USA\\ 3) Materials Sciences Division,
Lawrence Berkeley National Laboratory, Berkeley, California, 94720,
USA}
\date{\today}
\begin{abstract}

We study Fermi gases in a three-dimensional cubic optical lattice with five fermions per site, i.e. the s-band is completely filled and the $p$-band
with three-fold degeneracy is half filled. We show that, for repulsive interaction between fermions, the system will exhibit spin-$3/2$
antiferromagnetic order at low temperature. This conclusion is obtained both in strong interaction regime by strong coupling expansion, and in weak
interaction regime by Hatree-Fock mean-field theory with analysis of the Fermi surface nesting. We also show that, in the strongly correlated regime,
the N\'eel temperature for $p$-band antiferromagnetism is two to three orders of magnitudes higher than that of $s$-band, which is much more
promising to be attained in cold atom experiments.

\end{abstract}
\maketitle

\section{Introduction}

Studying strong correlations in lattice Fermi gases has now become an emerging forefront of cold atom physics, mainly because of the flexibility of
varying interactions and controlling filling number to access various quantum phases. One major effort pursued in many labs now is to use ultracold
Fermi gases to simulate Hubbard model, which exhibits various metallic, insulating and superconducting phases, with different magnetic behaviors
\cite{phase}. Despite of huge efforts extending over several decades, there are still lots of controversial issues on this model, which hopefully
cold atom experiments can shed light on.

The most solid conclusion of Hubbard model is drawn for a half-filled cubic lattice which says a spin-$1/2$ antiferromagnetic (AF) order exists for
all range of repulsive interaction. For weak interaction, the AF order arises from the nesting geometry of Fermi surface, while for strong
interaction, it is caused by the super-exchange interaction \cite{Auerbach,stringari,Georges,dmft-qmc}. Observing this AF order in cold atom
experiments, especially in strong interaction regime, is considered as a hallmark of seeing strong correlations in lattice Fermi gases and the
starting point for more ambitious goals along this direction.

On the other hand, in optical lattices physics can be much more richer than the single-band Hubbard model. When there are more
than two fermions at one lattice site, fermions will start to occupy excited bands like the $p$-band. Unlike in condensed matter
systems, there is no Jahn-Teller effect in optical lattices, so the three-fold orbital degeneracy of the $p$-band is well
maintained.

Recently there are increasing theoretical and experimental interests in studying bosons in the $p$-band, where a host of intriguing quantum phenomena
due to orbital degeneracy has been pointed out \cite{Bloch,Girvin,Congjun,XU,Zoo}. We expect the physics of $p$-band fermions will be more easier for
experimentally study, because the Pauli exclusion principle prevents the decay into the lower band. Thanks to the orbital degeneracy,  $p$-band Fermi
gases are expected to exhibit more diverse phases comparing to single-band Hubbard model. Nevertheless, the first step toward revealing these
exciting phases is to understand the unfrustrated spin order at half-filling. Here we consider a three-dimensional cubic lattice whose three
degenerate p-bands are half filled by three fermions per site. (The $s$-band is already completely filled by another two fermions and do not need to
be worried about.)

With the complexity in p-band model, even for half-filling there are issues not clear, for instance (i) whether the p-band
fermi-surface still has some nesting properties which give rise to a magnetic instability in weak interaction regime; (ii) with
multiple hopping channels in different orbitals and multiple choices of intermediate states, whether super-exchange processes can
still yield a simple Heisenberg-type model as in the s-band situation; and (iii) as there are various types of orders due to
orbital degrees of freedom, whether the strong and weak interaction regime share the same order or there is quantum phase
transition in between. These questions will be answered in this work. The main results of this paper can be summarized as:

({\bf I}) In strong interaction regime, at each isolated site the ground state of three fermions in p-orbitals are spin-$3/2$
states with four-fold degeneracy due to the Hund's rule. The coupling between two neighboring spins is caused by virtual hopping
of fermions, which gives rise to an isotropic spin-$3/2$ Heisenberg model:
\begin{equation}
{\bf H}_{J}=J_{\text{ex}}\sum_{\la ij\ra}\left(\vec{S}_i\vec{S}_j-\frac{1}{4}n_in_j\right)\label{HM}
\end{equation}
where $\vec{S}_i$ is an on-site spin-$3/2$ operator. $J_{\text{ex}}>0$ which means the ground state of this system has spin-$3/2$
antiferromagnetic order.

({\bf II}) In weak interaction regime, the Fermi surface of a half-filled $p$-band exhibits a perfect nesting symmetry, with the
nesting momentum $\vec{\pi}=(\pi,\pi,\pi)$. Mean-field analysis shows this nesting will induce antiferromagnetic order at
low-temperature, and the order parameter coincides with the $S=3/2$ spin order in strong interaction regime. We also show that
this order should be the only order parameter.

Based on ({\bf I}) and ({\bf II}), we anticipate that, despite the complexity with p-band orbital
degeneracy, the half-filled p-band Hubbard model exhibits only the spin-$3/2$ antiferromagnetic order for all range of
interaction strength, and the system undergoes a {\it crossover} from weak to strong interaction, a situation similar to the
half-filled single-band Hubbard model.

In the experiments of cold atoms in optical lattices, the major difficulty in achieving AF order in strongly correlated regime
comes from the fact that the N\'eel temperature is very low because of the ultra-small energy scale of super-exchange
interaction. In this work we show that in strong interaction regime the N\'eel temperature for $p$-band AF can reach a few
nano-Kelvin, which is two to three orders of magnitude higher than that of $s$-band. Because of this we suggest to study $p$-band
Fermi gas as a more realistic and promising route to reveal strongly correlated physics in optical lattices.

\begin{figure}[bp]
\includegraphics[angle=0,scale=0.37]
{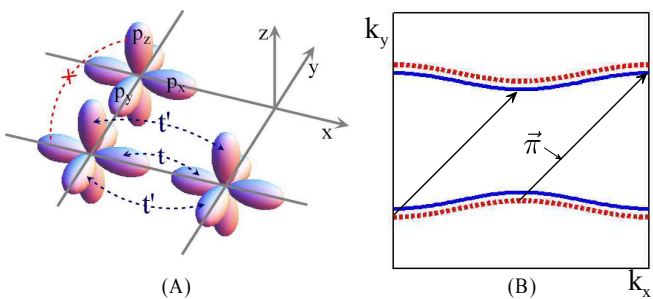} \caption{(color on-line) (A) Schematic of fermion hopping in $p$-band. For a fermion in $p_{x}$ orbital in $i$-site, its
hopping amplitude to $i+\hat{x}$ site is $t$, to $i+\hat{y}$-site and $i+\hat{z}$ site is $t^\prime(t'\ll t)$, and it can not hop
to $p_{y}$ (or $p_{z}$) orbital of its neighboring site. (B) Illustration of the $p$-band Fermi surface nesting at half-filling.
Red line and blue line are Fermi surfaces in $k_x-k_y$ plane for $k_z=\pi/3$ and $k_z=-2\pi/3$, respectively. These two Fermi
surfaces parallel with the nesting momentum $\vec{\pi}=(\pi,\pi,\pi)$(the solid black arrow). \label{model}}
\end{figure}

\section{ Model} The model for Fermi gases in optical lattices contains a nearest neighbor hopping ${\bf H}_{t}$ and an on-site
interaction ${\bf H}_{U}$. The hopping term ${\bf H}_{t}$ can be written as:
\begin{equation}
{\bf
H}_{t}=\sum\limits_{i,s}^{\alpha,\beta=x,y,z}T^{p_\alpha}_{\beta}c^\dag_{i,p_\alpha,s}c_{i+\hat{\beta},p_\alpha,s}+\text{h.c.},
\end{equation}
Since optical lattice is a separable potential in three dimensions ($V_0(\sin^2Kx+\sin^2Ky+\sin^2Kz)$), the hopping term conserves orbital index due
to the orthogonality of wannier wave functions, and the hopping amplitude is anisotropic
$T^{p_\alpha}_{\beta}=t\delta_{\alpha\beta}-t^\prime(1-\delta_{\alpha\beta})$ with $t,t'>0$. And there is also no next nestest neighbor hopping
\cite{Georges}. Fig. \ref{model}(A) is a schematic of hoppings in the $p$-band. $t$ is much larger than $t^\prime$, for instance, for an optical
lattice with the depth $V_0=11E_{\text{R}}$  ($E_{\text{R}}=\hbar^2K^2/(2m)$ is the photon recoil energy), $t\approx 20t^\prime$. For ultra-cold
neutral gases, the interaction is dominated by short-range $s$-wave repulsion $V({\bf r_{12}})=(4\pi\hbar^2 a_{\text{s}}/m)\delta({\bf r_1-r_2})$,
which takes place between unlike spins only \cite{stringari}. The on-site interaction ${\bf H}_{U}$ then has the form
\begin{eqnarray}
{\bf H}_{U}=\sum\limits_i\left\{U\sum_{\alpha}n_{i,p_\alpha,\uparrow}n_{i,p_\alpha,\downarrow}+W\sum\limits_{\alpha\neq \beta}(n_{i,p_\alpha,\uparrow}n_{i,p_\beta,\downarrow}\right.\nonumber\\
\left.+c^\dag_{i,p_\alpha,\uparrow}c^\dag_{i,p_\beta,\downarrow}c_{i,p_\alpha\downarrow}c_{i,p_\beta,\uparrow}+c^\dag_{i,p_\alpha,\uparrow}c^\dag_{i,p_\alpha,\downarrow}c_{i,p_\beta,\downarrow}c_{i,p_\beta,\uparrow})\right\},\\
U=\frac{4\pi \hbar^2a_{\text{s}}}{m}\int d^3{\bf r}|w_{p_\alpha}({\bf r})|^4 \ \ \ \  \ \  \ \   \   \ \   \ \ \   \   \   \    \   \   \   \   \   \   \    \   \     \  \    \    \       \  \  \  \   \\
W=\frac{4\pi \hbar^2a_{\text{s}}}{m}\int d^3{\bf r}|w_{p_\alpha}({\bf r})|^2|w_{p_\beta}({\bf r})|^2\ \ \ \ \  \ \ \ \ \  \  \ \
\ \ \ \ \ \ \ \ \
\end{eqnarray}
where $w_{p_\alpha}$ is the $p_\alpha$-orbital wannier wave function. There are totally four terms due to $s$-wave interaction. The first term is the interaction within the same orbital, the second term is the interaction between unlike orbitals, the third term represents the spin exchange between unlike orbitals and the fourth term represents two fermions hopping processes from one orbital to another. Usually $U$ is three times larger than $W$.

\section{ Strong Interaction Regime}  When the on-site interaction is dominative, we shall first ignore the hopping term and calculate
all the eigenstates of an isolated site. The results for three, four and two fermions per site are summarized in following Table \ref{N3}, \ref{N4}
and \ref{N2}. As one can see from Table \ref{N3}, the low-energy Hilbert space is constituted by four spin-$3/2$ states. Because these spin states
are symmetric superpositions of three atoms, their spatial wave functions have to be antisymmetric and the interactions are completely quenched. In
this limit, the system is a spin-3/2 Mott insulator \cite{highspin} .

\begin{table}[h]
\begin{center}
\begin{tabular}{|c|c|c|}
\hline
$E=0 (S=3/2)$ & $E=3W (S=1/2)$&$E\geq U$\\
\hline
$ |\uparrow \uparrow \uparrow\rangle$ &$(|\uparrow \uparrow \downarrow\rangle-|\uparrow \downarrow \uparrow\rangle)/\sqrt{2}$&\text{double} \\
$|\downarrow \downarrow \downarrow\rangle$ & $(|\downarrow \uparrow \uparrow\rangle-|\uparrow \downarrow \uparrow\rangle)/\sqrt{2}$ & \text{occupied} \\
$(|\downarrow \uparrow\uparrow\rangle+|\uparrow \downarrow \uparrow\rangle+|\uparrow \uparrow \downarrow)/\sqrt{3}$& $(|\downarrow \downarrow\uparrow\rangle-|\downarrow \uparrow \downarrow\rangle)/\sqrt{2}$ &states \\
$(|\downarrow \downarrow\uparrow\rangle+|\uparrow \downarrow \downarrow\rangle+|\downarrow \uparrow \downarrow)/\sqrt{3}$& $(|\uparrow \downarrow \downarrow\rangle-|\downarrow \uparrow\downarrow\rangle)/\sqrt{2}$ &\\
\hline
\end{tabular}
\caption{Eigenstates and eigenenergies of three fermions at an isolated site. In these three tables, we define that, for example,
$|d\uparrow0\rangle$ means $p_x$ is double-occupied, $p_y$ is occupied by a spin-$\uparrow$ fermion, and $p_z$ is empty. \label{N3}}
\end{center}
\label{default}
\end{table}%
\begin{table}[h]
\begin{center}
\begin{tabular}{|c|c|}
\hline
$E=U+2W (S=1)$ & $E=U+4W (S=0)$\\
\hline
$ |d \uparrow \uparrow\rangle$, $ |\uparrow d \uparrow\rangle$, $ |\uparrow \uparrow d\rangle$, $ |d \downarrow \downarrow\rangle$,  &$(|d \uparrow \downarrow\rangle-|d \downarrow \uparrow\rangle)/\sqrt{2}$ \\
$ |\downarrow d \downarrow\rangle$, $ |\downarrow \downarrow d\rangle$, $(|d \uparrow \downarrow\rangle+|d \downarrow \uparrow\rangle)/\sqrt{2}$& $(| \uparrow d \downarrow\rangle-| \downarrow d \uparrow\rangle)/\sqrt{2}$  \\
$(| \uparrow d \downarrow\rangle+| \downarrow d \uparrow\rangle)/\sqrt{2}$  & $(| \uparrow \downarrow d\rangle-| \downarrow\uparrow d\rangle)/\sqrt{2}$  \\
$(| \uparrow \downarrow d\rangle+| \downarrow\uparrow d\rangle)/\sqrt{2}$   &   \\
\hline
\end{tabular}
\caption{Eigenstates and eigenenergies of four fermions at an isolated site. \label{N4}}
\end{center}
\label{default}
\end{table}%
\begin{table}[h]
\begin{center}
\begin{tabular}{|c|c|c|}
\hline
$E=0 (S=1)$ & $E=2W (S=0)$&$E\geq U-W$\\
\hline
$ |0 \uparrow \uparrow\rangle$, $ |\uparrow 0 \uparrow\rangle$, $ |\uparrow \uparrow 0\rangle$, &$(|0 \uparrow \downarrow\rangle-|0 \downarrow \uparrow\rangle)/\sqrt{2}$ & double \\
$ |0 \downarrow \downarrow\rangle$, $ |\downarrow 0 \downarrow\rangle$, $ |\downarrow \downarrow 0\rangle$ & $(| \uparrow 0 \downarrow\rangle-| \downarrow 0 \uparrow\rangle)/\sqrt{2}$   & occupied \\
$(|0 \uparrow \downarrow\rangle+|0 \downarrow \uparrow\rangle)/\sqrt{2}$ & $(| \uparrow \downarrow 0\rangle-| \downarrow\uparrow 0\rangle)/\sqrt{2}$  & states \\
 $(| \uparrow 0 \downarrow\rangle+| \downarrow 0 \uparrow\rangle)/\sqrt{2}$
&  &\\
$(| \uparrow \downarrow 0\rangle+| \downarrow\uparrow 0\rangle)/\sqrt{2}$   & & \\
\hline
\end{tabular}
\caption{Eigenstates and eigenenergies of two fermions at an isolated site.\label{N2}}
\end{center}
\label{default}
\end{table}%

The coupling of spins between neighboring sites can be obtained by taking the hopping term as the second-order perturbation with the Brillioun-Wigner
approximation\cite{Auerbach}:
\begin{equation}
{\bf H}_{J}=-\sum_{e}\mathcal{P}_{G}{\bf H}_{t}\mathcal{P}_{e}\frac{1}{{\bf H}_{U}}\mathcal{P}_{e}{\bf H}_t \mathcal{P}_{G},
\end{equation}
where ${\mathcal P}_{e}$ means the projection into excited Hilbert space, and ${\mathcal P}_{G}$ means the projection into $S=3/2$ Hilbert space. To
proceed we make following observations: (i) The virtual hopping processes only take place within the same orbital, otherwise the final state will
have double occupancy, which should be projected out. For instance, if a $p_x$ fermion hops from $i$-site to $j$-site, it must be followed by a $p_x$
fermion hops from $j$-site to $i$-site. (ii) By using Table \ref{N3}, \ref{N4} and \ref{N2} one can show that only one type of excited states can be
connected to the unperturbed ground states via hopping, which is the lowest energy states of four fermions ($S=1$ and $E=U+2W$) in one site and the
lowest energy states of two fermions ($S=1$ and $E=0$) in its neighborhood \cite{vitual-hopping}. So the intermediate states always has energy
$U+2W$.

Because of (i-ii), ${\bf H}_{J}$ on the link between $i$ and
$i+\hat{x}$ site reads
\begin{eqnarray}
\sum\limits_{p_\alpha}\frac{2|T_{x}^{p_\alpha}|^2}{U+2W}\mathcal{P}_{G}\left(\sum\limits_{ss^\prime}c^\dag_{i,p_\alpha,s}c_{i,p_\alpha,s^\prime}c^\dag_{i+\hat{x},p_\alpha,s^\prime}c_{i+\hat{x},p_\alpha,s}\right)\mathcal{P}_{G}\nonumber.
\end{eqnarray}
Note that
\begin{eqnarray}
c^\dag_{i,p_\alpha,s}c_{i,p_\alpha,s^\prime}=\frac{1}{3}\left(\sum_{p_\beta}c^\dag_{i,p_\beta,s}c_{i,p_\beta,s^\prime}+\sum_{p_\beta}\Xi^{s,s^\prime}_{i,p_\alpha,p_\beta}\right),\nonumber
\end{eqnarray}
where
\begin{equation}\Xi^{s,s^\prime}_{i,p_\alpha,p_\beta}=c^\dag_{i,p_\alpha,s}c_{i,p_\alpha,s^{\prime}}-c^\dag_{i,p_\beta,s}c_{i,
p_\beta,s^\prime},
\end{equation}
and with the help of Table \ref{N3}, one can show that
\begin{eqnarray}
&&\mathcal{P}_{G}\Xi^{s,s^\prime}_{i,p_\alpha,p_\beta}\mathcal{P}_{G}=0\\
&&\mathcal{P}_{G}\sum_{p_\beta}c^\dag_{i,p_\beta,s}c_{i,p_\beta,s^\prime}\mathcal{P}_{G}=\sum_{p_\beta}c^\dag_{i,p_\beta,s}c_{i,p_\beta,s^\prime}.
\end{eqnarray}
Then $H_{J}$ on this link can be written as
\begin{eqnarray}
\sum\limits_{p_\alpha}\frac{2|T_{x}^{p_\alpha}|^2}{9(U+2W)}\sum\limits_{ss^\prime}\sum_{p_\beta}c^\dag_{i,p_\beta,s}c_{i,p_\beta,s^\prime}\sum_{p_\beta^\prime}c^\dag_{i,p_\beta^\prime,s}c_{i,p_\beta^\prime,s^\prime}\nonumber\\
={1\over2}J_{\text{ex}}\sum\limits_{p_\alpha,p_\beta}\sum\limits_{s,s^\prime}\left(c^\dag_{i,p_\alpha,s}c_{i,p_\alpha,s^\prime}\right)\left(c^\dag_{i+\hat{x},
p_\beta, s^\prime}c_{i+\hat{x},p_\beta,s}\right)\ \ \nonumber
\end{eqnarray}
where
\begin{equation}
J_{\text{ex}}=4(t^2+2t^{\prime 2})/(9U+18W).
\end{equation} We now introduce $\vec{S}_i$ as
\begin{equation}
\vec{S}_i=\frac{1}{2}\sum\limits_{p_\alpha}\sum\limits_{ss^\prime}c^\dag_{i,p_\alpha,s}\vec{\sigma}_{ss^\prime}c_{i,p_\alpha,s^\prime}\label{S3/2}
\end{equation}
and $\vec{\sigma}$ are the Pauli matrices. Restricted in the low-energy Hilbert space, $\vec{S}_i$ acts as $S=3/2$ spin operator. Hence, we have
shown the first main result that

\textsf{In the strong interaction regime, the super-exchange processes induce an isotropic spin-$3/2$ Heisenberg model \begin{equation}
{\bf H}_{J}=J_{\text{ex}}\sum_{\la ij\ra}\left(\vec{S}_i\vec{S}_j-\frac{1}{4}n_in_j\right)\label{HM}
\end{equation}
which describes the low-energy spin dynamics of $p$-band Mott insulator. $J_{\text{ex}}>0$. Below N\'eel temperature, it will
give rise to an antiferromangetic long range order of $\la S_i^+\ra=(-1)^i|\la S_i^+\ra|\neq0$}

\section{Weak Interaction Regime} In this regime we use usual Hartree-Fock mean-field theory, for which the Hamiltonian is rewritten into momentum
space. The hopping term can be written as:
\begin{equation}
{\bf H}_t=\sum\limits_{{\bf k},\alpha}\epsilon_{\bf k}^{p_\alpha}c^\dag_{p_\alpha,{\bf k}}c_{p_\alpha,{\bf k}},
\end{equation}
where
\begin{equation}\epsilon_{\bf k}^{p_\alpha}=2t\cos k_\alpha-2t^\prime\sum_{\beta\neq\alpha}\cos k_\beta-\mu.
\end{equation} For half-filling, $\mu=0$, and the Fermi surface for one of the $p$-bands is plotted in Fig. \ref{model}(B).

For performing mean-field decomposition, we
reconstruct the interaction term ${\bf H}_{U}$ as ${\bf H}^0_U+{\bf H}^\prime_U$, where
\begin{eqnarray}
&&{\bf H}^0_U=-V_1\sum\limits_{{\bf k,k^\prime},\bf q}\hat{\Delta}^\dag_{{\bf k,q}}\hat{\Delta}_{{\bf k^\prime,q}},\\
&&{\bf H}^\prime_{U}=\sum\limits^{\alpha\neq\beta}_{{\bf k,k^\prime},q}\left(-V_2\hat{\Gamma}^{\alpha\beta\dag}_{{\bf
k,q}}\hat{\Gamma}^{\alpha\beta}_{{\bf k^\prime,q}}-2W\hat{\Sigma}^{\alpha\beta\dag}_{{\bf k,q}}\hat{\Sigma}^{\alpha\beta}_{{\bf
k^\prime,q}}\right)
\end{eqnarray}
with $V_1=3(U+2W)$ and $V_2=2(U-W)/3$. Here we introduce
\begin{eqnarray}\hat{\Delta}^\dag_{{\bf k,q}}=\sum_{p_\alpha}S^{{\bf
q}+}_{p_\alpha,p_\alpha,{\bf k}}/3,
\end{eqnarray} and \begin{eqnarray}\hat{\Gamma}^{\alpha\beta\dag}_{{\bf k,q}}=(S^{{\bf q}+}_{p_\alpha,p_\alpha,{\bf
k}}-S^{{\bf q}+}_{p_\beta,p_\beta,{\bf k}})/2,\\
\hat{\Sigma}^{\alpha\beta\dag}_{{\bf k,q}}=(S^{{\bf q}+}_{p_\alpha,p_\beta,{\bf k}}+S^{{\bf q}+}_{p_\beta,p_\alpha,{\bf k}})/2,\end{eqnarray} where
$S^{{\bf q}+}_{p_\alpha,p_\beta,{\bf k}}= c^\dag_{{\bf k},p_\alpha,\uparrow}c_{{\bf k+q}, p_\beta,\downarrow}$ is the spin lifting operator. As
illustrated in Fig. \ref{model}(B), each of three $p$-band Fermi-surfaces exhibits nesting symmetry with the nesting momentum
$\vec{\pi}=(\pi,\pi,\pi)$ at half-filling. It means that for each ${\bf k}$ at Fermi surface, ${\bf k+\pi}$ also locates at Fermi surface, therefore,
the maximum contribution to ${\bf H}_U$ comes from ${\bf q}=\vec {\pi}$. For this reason we only consider ${\bf q} =\vec{\pi}$ component in the
mean-field approximation, and we introduce three order parameters which are $\Delta=V_1\sum_{{\bf k}}\langle \hat{\Delta}_{{\bf k,\pi}}\rangle$,
$\Gamma=V_2\sum_{{\bf k}}\langle \hat{\Gamma}^{\alpha\beta}_{{\bf k,\pi}}\rangle/3$ and $\Sigma=2W\sum_{{\bf k}}\langle
\hat{\Sigma}^{\alpha\beta}_{{\bf k},\pi}\rangle$, where $\Gamma$ and $\Sigma$ are independent of index $\alpha$, $\beta$ because of the permutation
symmetry between orbitals. Note that $\Delta$ is the same spin order as in strong interaction regime, we separate the mean-field Hamiltonian into two
parts as ${\bf H}_{\text{MF}}={\bf H}_0+{\bf H}^\prime$ where
\begin{eqnarray}
{\bf H}_0=\sum\limits_{{\bf k}}\left[\sum\limits_{p_\alpha,\sigma}\epsilon_{{\bf k}}^{p_\alpha}c^\dag_{{\bf
k},p_\alpha,\sigma}c_{{\bf k},p_\alpha,\sigma}+(\Delta\hat{\Delta}^\dag_{\bf k}+\text{h.c.})\right]+\frac{\Delta^2}{V_1}\nonumber
\end{eqnarray}
and
\begin{eqnarray}
{\bf H}^\prime=\sum\limits^{\alpha\neq\beta}_{{\bf k}}\left[\Sigma\hat{\Sigma}^{\alpha\beta\dag}_{{\bf
k}}+\text{h.c.}\right]+\frac{\Gamma^2}{V_2}+\frac{\Sigma^2}{2W}\nonumber
\end{eqnarray}
Note that ${\bf H}^\prime$ does not contain $\hat{\Gamma}$-terms because they all cancel out due to the permutation symmetry of
three orbitals, therefore $\Gamma=0$.

Because the three orbitals are now separable in ${\bf H}_0$, it can be easily diagnolized by Bogoliubov
transformation
\begin{equation}
{\bf H}_0=\sum\limits_{{\bf k},\alpha}\xi_{{\bf k}}^{p_\alpha}(\gamma^\dag_{{\bf k},p_\alpha}\gamma_{{\bf
k},p_\alpha}+\eta^\dag_{{\bf k},p_\alpha}\eta_{{\bf k},p_\alpha}-1)+\frac{\Delta^2}{V_1},
\end{equation}
where
\begin{equation}\xi_{{\bf k}}^{p_\alpha}=\sqrt{(\epsilon_{\bf k}^{p_\alpha}-\epsilon_{{\bf k}+\vec{\pi}}^{p_\alpha})^2/4+(\Delta/3)^2},
\end{equation}
and $\gamma_{{\bf k},p_\alpha}$ and $\eta_{{\bf k},p_\alpha}$ are quasi-particle operators\cite{gamma-eta}. The equation for $\Delta$
is
\begin{equation}
\frac{1}{V_1}=\sum\limits_{{\bf k},p_\alpha}\frac{\tanh(\xi_{\bf k}^{p_\alpha}\beta/2)}{\xi_{\bf k}^{p_\alpha}},\label{MF}
\end{equation}
the solution of which yields non-zero value of $\Delta$ and the critical temperature for having non-zero $\Delta$.

Next we shall examine whether there will be symmetry breaking of $\Sigma$ also, for which we treat ${\bf H}^\prime$ as perturbation on top of ${\bf
H}_0$, and the energy can be expanded in powers of $\Sigma$ as $ E_{\text{MF}}=E_0(\Delta)+\chi(\Delta)\Sigma^2+O(\Sigma^4) $. So whether there is an
instability for developing finite $\Sigma$ depends on the sign of $\chi$. Straightforward calculation shows that
\begin{equation}
\chi(\Delta)=-\sum\limits_{{\bf k},{\alpha\neq \beta}}\frac{|\tau^{\alpha\beta}_{{\bf k}}|^2}{2(\xi_{\bf k}^{p_\alpha}+\xi_{{\bf
k}+\vec{\pi}}^{p_\beta})}+\sum\limits_{{\bf k},\alpha}\frac{V_1}{2W}\frac{1}{\xi_{\bf k}^{p_\alpha}},\label{W}
\end{equation}
where $\tau^{\alpha\beta}_{{\bf k}}=v_{{\bf k},p_\alpha}v_{{\bf k}+\vec{\pi},p_\beta}-u_{{\bf k},p_\alpha}u_{{\bf k}+\vec{\pi},p_\beta}$
\cite{gamma-eta}. We numerically compute Eq. (\ref{W}) and find that it is always positive, which indicates that there will be no symmetry breaking
of $\Sigma$. Mathematically, it is because $\xi_{{\bf k},p_\alpha}\neq\xi_{{\bf k}+\vec{\pi},p_\beta}$ except for a few specific points in momentum
space. In the limit of $\Delta\rightarrow 0$, the first term in the r.h.s. of Eq. (\ref{W}) diverges only around those a few points, while the second
term diverges around the whole Fermi surface, hence the second term will always be dominative and the whole summation is positive.  Physically, it is
because the AF instability crucially relies on the nesting of Fermi surface, and the fact that there is no nesting between two different $p$-bands
prevents inter-orbital spin order. Hence we have reached the second main result that

\textsf{In the weak interaction regime, the only spin order at low-temperature is
$\Delta=\langle\sum_{{\bf k},\alpha} c^\dag_{{\bf k},p_\alpha\uparrow} c_{{\bf
k+\vec{\pi}},p_\alpha,\downarrow}\rangle\neq 0$ and this order coincides with that in strong interaction regime.}

\section{N\'eel Temperature}
\begin{figure}[tbp]
\includegraphics[angle=0,scale=0.45]
{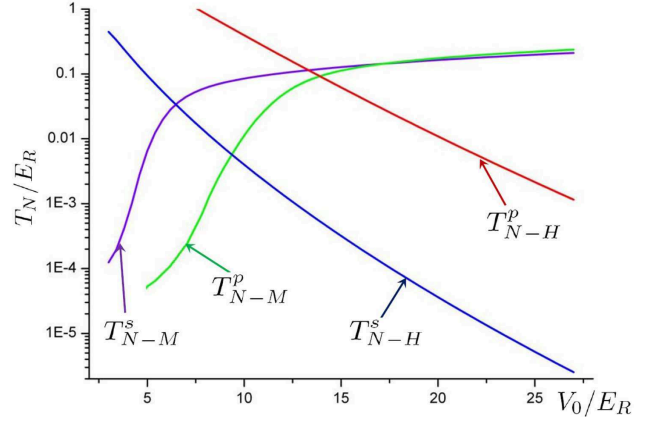} \caption{(color on-line) N\'eel temperature $T_N$ as a function of optical lattice depth $V_0$. Four curves are
$p$-band $T_N$ from Heisenberg model $T^p_{N-H}$, from mean-field theory $T^p_{N-M}$; $s$-band $T_N$ from Heisenberg model
$T^s_{N-H}$, from mean-field theory $T^s_{N-M}$, respectively.   For parameters, we use $^{40}K$ in optical lattices as an
example \cite{Essilinger}. The lattice spacing $\lambda=1064nm$, $K=\pi/\lambda$ and $a_{\text{s}}=15nm$. The energy units is
taken as $E_{\text{R}}$, which corresponds to $0.21\mu K$.\label{Tc}}
\end{figure}

Here we first calculate the wannier functions in an optical lattice potential and use them to deduce the parameters $t$, $t^\prime$, $U$ and $W$ in
our model. Then, we calculate the N\'eel temperature given by Heisenberg model of Eq.(\ref{HM}) $T_{N-H}$, which is
$1.276J_{\text{ex}}S(S+1)$,\cite{Georges} and also calculate $T_{N-M}$ given by mean-field theory by solving equation (\ref{MF}) for the onset of
non-zero $\Delta$. We display our calculation of these two $T_N$ for both half-filled $p$-band and $s$-band in Fig. \ref{Tc}.  $T_{N-H}$  decreases
as the increase of the lattice height while $T_{N-M}$ increases as the increase of the lattice height.

For small $V_0$, $T_{\text{N-M}}\ll T_{\text{N-H}}$, the AF order is caused by mean-field effect, while for large $V_0$ where $T_{\text{N-H}}\ll
T_{\text{N-M}}$, the AF order is driven by super-exchange processes and the system is in strongly correlated regime. The regime where
$T_{\text{N-M}}$ and $T_{\text{N-H}}$ cross is the crossover regime from weak to strong interactions, where both two methods will fail, and more
advanced numerical technique like DMFT and QMC is required in order to give accurate N\'eel temperature \cite{dmft-qmc}. We will leave this for
further investigation. However, the key point is that the s-band $T_{\text{N-H}}$ is already of the order of $0.1 \text{nK}$ when the lattice height
enters the strongly correlated regime of great experimental interests, and it decreases to below $10^{-2}\text{nK}$ rapidly as $V_0$ increases, while
the p-band $T_{\text{N-H}}$ remains the order of $1\text{nK}$ even when $V_0$ increases to $\sim 20 E_{\text{R}}$. It is simply because the wannier
wave function in $p$-band is much more extended, its hopping amplitude $t$ is one to two orders larger than that of $s$-band, which consequently
leads to much larger $J_{\text{ex}}$.

In additional to higher N\'eel temperature, the spin-$3/2$ of $p$-band AF state could have a better signal-to-noise radio than $s$-band spin-$1/2$ AF. All these provide great advantage for experimental study of
quantum AF and its related phenomena due to super-exchange. Besides, we leave the effect of trapping potential to future investigation which may introduce shell structure in density profile and the finite size effect into the problem.

We hope that this work will open up a new route for studying strongly
correlated Fermi gases in optical lattices, and be bases for future efforts along this direction.

{\it Acknowledgment}: We thank Jason Ho for critical reading our manuscript and valuable comments, we thank Zheng-Yu Weng for
helpful discussion, and Xi Dai for fruitful discussion on Fermi surface nesting. HZ would like to thank Kavli Institute for
Theoretical Physics China for hospitality. This work is supported by NSFC under grant No. 10547002.


\begin{thebibliography}{99}
\bibitem{vitual-hopping} It is because the lowest energy states in each table are all constructive superposition of different states, while other states are destructive superposition.
\end{thebibliography}
\end{document}